\documentclass[aps,twocolumn,superscriptaddress]{revtex4}
\usepackage{amsmath}
\usepackage{amsfonts}
\usepackage{amssymb}
\usepackage{epsfig}

\makeatletter
\newcommand*{\rightharpoonupfill@}{%
  \arrowfill@\relbar\relbar\rightharpoonup
}

\newcommand*{\leftharpoondownfill@}{%
  \arrowfill@\leftharpoondown\relbar\relbar
}

\newcommand{\xrightleftharpoons}[2][]{%
  \ensuremath{%
    \mathrel{%
      \settoheight{\dimen@}{\raise 2pt\hbox{$\rightharpoonup$}}%
      \setlength{\dimen@}{-\dimen@}%
      \edef\CA@temp{\the\dimen@}%
      \settoheight\dimen@{$\rightleftharpoons$}%
      \addtolength{\dimen@}{\CA@temp}%
      \raisebox{\dimen@}{%
        \rlap{%
          \raisebox{2pt}{%
            $%
            \ext@arrow 0359\rightharpoonupfill@{\hphantom{#1}}{#2}%
            $%
          }%
        }%
        \hbox{%
          $%
          \ext@arrow 3095\leftharpoondownfill@{#1}{\hphantom{#2}}%
          $%
        }%
      }%
    }%
  }%
}
\makeatother

\newcommand{\harp}{\xrightleftharpoons}
\newcommand{\xa}{\xrightarrow[]}

\newcommand{\s}{\sigma}

\renewcommand{\r}{\rho}

\newcommand{\beq}{\begin{equation}}
\newcommand{\eeq}{\end{equation}}
\newcommand{\bea}{\begin{eqnarray}}
\newcommand{\eea}{\end{eqnarray}}
\newcommand{\bal}{\begin{align}}
\newcommand{\eal}{\end{align}}

\usepackage{dsfont}

\begin{document}
\title{Geometry and flexibility of optimal catalysts in a minimal elastic network model}

\author{Olivier Rivoire} 

\affiliation{Center for Interdisciplinary Research in Biology (CIRB), Coll\`ege de France, CNRS, INSERM, PSL Research University, Paris, France}


\begin{abstract}
We have a general knowledge of the principles by which catalysts accelerate the rate of chemical reactions but no precise understanding of the geometrical and physical constraints to which their design is subject. To analyze these constraints, we introduce a minimal model of catalysis based on elastic networks where the implications of the geometry and flexibility of a catalyst can be studied systematically. The model demonstrates the relevance and limitations of the principle of transition-state stabilization: optimal catalysts are found to have a geometry complementary to the transition state but a degree of flexibility that non-trivially depends on the parameters of the reaction as well as on external parameters such as the concentrations of reactants and products. The results illustrate how simple physical models can provide valuable insights on the design of catalysts.
\end{abstract}

\maketitle

Catalysts, which increase the rate of chemical reactions without being part of their products, are essential to biological processes as well as to the industrial production of most chemicals. We have a general theory of catalysis, transition-state theory~\cite{Eyring:1935te,Evans:1935dg}, and detailed knowledge of the mechanisms by which many catalysts operate, in particular enzymes~\cite{fersht1999structure}. We also have an increasing capacity to model and numerically simulate catalytic processes at an atomic level~\cite{karplus2002molecular}. Yet, basic questions pertaining to the existence of fundamental geometrical and physical constraints to catalysis are still the object of speculations: To what extent does efficient catalysis require catalysts to be rigid?~\cite{Kraut:1988th} Or thermally stable?~\cite{karshikoff2015rigidity} Does it impose a minimal size on catalysts?~\cite{Srere:1984wf}  Is catalysis subject to a general rate-accuracy trade-off?~\cite{tawfik2014accuracy}

Answers to such questions would help us uncovering the design principles of natural enzymes~\cite{davidi2018bird}, directing the experimental evolution of novel enzymes~\cite{goldsmith2012directed}, and clarifying the conditions under which life can emerge~\cite{walker2017re}.

Missing is a theoretical framework that is sufficiently elaborate to account for geometric and physical constraints, yet sufficiently simple to allow for a systematic comparison of varied geometries and physical designs. For this purpose, the low-dimensional phase-space formulation of transition-state theory is too abstract, as it does not refer explicitly to the spatial architecture of catalysts. The atom-level description of models studied by molecular dynamic simulations is, on the other hand, too detailed, as it prohibits computational exploration of a large number of architectures. 

An alternative lies in the simplified physical models developed to study properties of proteins other than catalysis, notably folding~\cite{Pande:1997cp}, binding~\cite{Miller:1997es} and allostery~\cite{eckmann2019colloquium}. Particularly insightful are elastic network models, which approximate molecules by a network of beads interacting through elastic springs~\cite{Chennubhotla:2005bb}. In their different guises, these models have provided conceptual and quantitative insights into several features of proteins, including thermal fluctuations~\cite{bahar1997direct}, conformational changes~\cite{tama2001conformational}, unfolding kinetics~\cite{Dietz:2008gj,Srivastava:2013hy}, specificity~\cite{Savir:2007bd,rivoire2018minimal} and allostery~\cite{McLeish:2013ij,yan2017architecture}. 

Here we propose to adapt the framework of elastic network models to study catalysis. We illustrate this proposal by defining and solving a one-dimensional model of catalysis. Our model may be viewed as a reformulation and systematic analysis of a model of strain-induced catalysis first suggested by Haldane~\cite{haldane1930enzymes} and later partly formalized by Gavish~\cite{Gavish:1978vr,Gavish86,Bustamante:2004fo}. While deliberately minimal, the model addresses a key design challenge: an efficient catalyst must stabilize the transition state of the reaction to accelerate it but also bind to the reactant and release the product. These conflicting demands lead to non-trivial constraints on flexibility, which our model recapitulates. The model also demonstrates how the optimal design of a catalyst depends, beyond the mechanisms of the reaction, on the conditions under which catalysis occurs. Our analysis is limited to one dimension but the model is straightforward to extend, if not to solve, in two or three dimensions. Our approach thus complements other bottom-up studies of catalysis~\cite{ Gavish:1978vr,Zeravcic:cm} towards a better understanding of the geometrical and physical constraints to which proficient catalysts are subject.

\section{General framework}

Analyzing the physical and geometrical constraints to efficient catalysis requires a physical model that specifies the range of designs to be examined and a criterion to quantify catalytic efficiency. Our choices in defining such a model are guided by a principle of simplicity, the goal being to obtain a physically coherent framework where a large number of different architectures can effectively be explored and compared.

\subsection{Physical model}

Elastic network models are one of the simplest physical models where geometry, strain and energy can be related. They consist of beads interacting through elastic springs and have been extensively used to study the internal motions of proteins~\cite{Chennubhotla:2005bb}. Each spring is characterized by two parameters, a spring constant and a free length. Varying the number of beads and the parameters of the springs that connect them allows for the sampling of a large number of designs, including networks approximating three-dimensional protein structures~\cite{Chennubhotla:2005bb}. Here, we propose to describe not only a catalyst, but also its substrate and their interaction within a common elastic network model. To this end, we assume that each spring has a maximal extension above which it breaks and below which it reforms. More precisely, each spring contributes to the total energy by $k(|x|-l)^2/2-k(z-l)^2/2$ if the extension $x$ satisfies $|x|<z$ and 0 if $|x|>z$, where $k>0$ is the spring constant, $l>0$ the free length and $z>l$ the maximal extension. When the beads are subject to Brownian motion, which accounts for their interaction with a solvent, bonds may thus break or form as a result of thermal fluctuations.

\begin{figure}[t]
\begin{center}
\includegraphics[width=.9\linewidth]{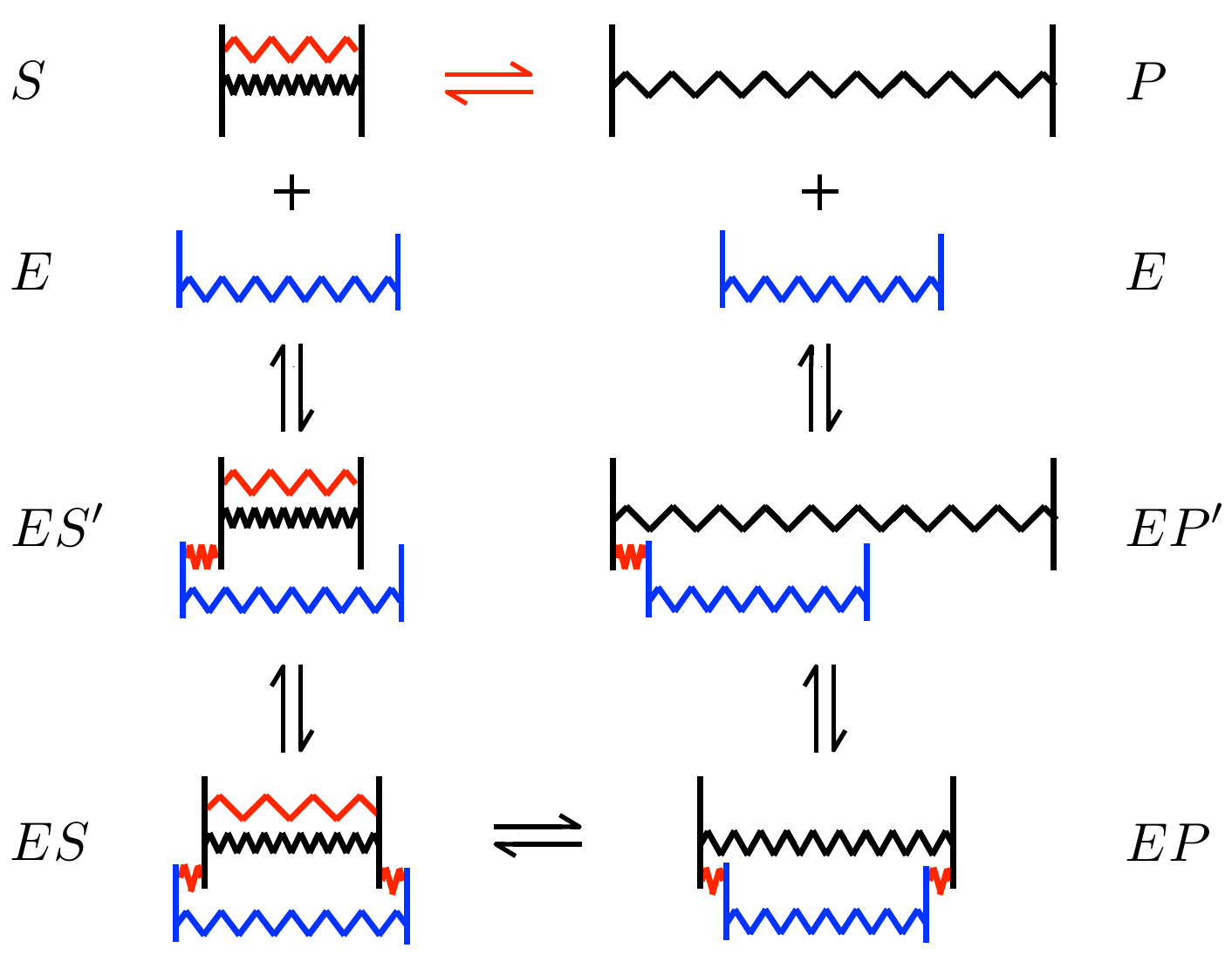}
\caption{Elastic network model of catalysis -- The reaction $S\xrightleftharpoons[]{} P$ is defined on the top. The reactant $S$ consists of two beads connected by two springs (here represented by vertical lines). One spring (in red) breaks when its extension exceeds a threshold, which results in the product $P$. The system is subject to thermal fluctuations and the reaction may thus occur spontaneously. A catalyst $E$ (in blue) similarly consists of two beads connected by a spring. Each bead of the catalyst can interact with one bead of the substrate through a breakable spring (in red) that forms when the distance between the two beads is below a threshold and breaks when their distance is above this same threshold. Six non-equivalent states can be distinguished, $S+E$, $ES'$, $ES$, $EP$, $EP'$ and $P+E$, depending on whether each type of breakable spring is broken or not.\label{fig:scheme}}
\end{center} 
\end{figure}

The rupture of a bond between two beads defines an elementary chemical reaction. To have a single product as well as a single reactant, we consider a case where this rupture does not compromise the connectivity of the substrate. This is achieved by assuming that a second unbreakable bond (with infinite maximal extension) links the two beads: the presence of the two springs then defines the reactant $S$ while the absence of the breakable spring defines the product $P$ (Fig.~\ref{fig:scheme}, top line).

In this framework, the simplest catalyst also consists of just two beads joined by a single unbreakable spring. To describe its interaction with the substrate, either in the form of the reactant $S$ or the product $P$, we assume that each bead of the catalyst can interact through a breakable spring with one, and only one, of the beads of the substrate (Fig.~\ref{fig:scheme}). 

In total, our elastic network model thus comprises four beads and five springs, three of which being effectively absent if their extension exceeds a given threshold.  Assuming the breakable springs to have a vanishing free  length, the model is then specified by 8 parameters (Table~\ref{tab:para}).

\begin{table}
\begin{center}
\begin{tabular}{|c|c|c|c|}
  \hline
 & spring  & free & maximal \\
 & constant & length & extension\\
  \hline
substrate scissile bond  & $k_a$ & $0$ & $z_a$\\
 \hline
substrate non-scissile bond & $k_r$ & $l_r$ & $\infty$\\
 \hline
catalyst internal bond & $k_e$ & $l_e$ & $\infty$\\
 \hline
substrate-catalyst interaction & $k_i$ & $0$ & $z_i$\\
 \hline
\end{tabular}
\end{center}
\caption{Eight parameters of the elastic network model --  Each bond has three parameters: a spring constant $k$, a free length $l$ that defines an elastic interaction and a maximal extension $z>l$ beyond which this interaction is no longer present. The substrate consists of two beads connected by two bonds, one scissile ($z_a<\infty$) and the other not ($z_r=\infty$). The catalyst consists of two beads connected by a single bond. The interaction between the beads of the substrate and those of the substrate are described by breakable springs. The free lengths of breakable springs is taken to be zero.\label{tab:para}}
\end{table}

\subsection{Criteria for catalytic efficiency}

There is no intrinsically optimal catalyst. Depending on the set-up, and not just the reaction to be catalyzed, different criteria are relevant for scoring catalytic activity. Optimizing these different criteria generally leads to different optimal designs. 

Consider for instance a measure of catalytic efficiency commonly adopted in enzymology, the ratio $k^+_{\rm cat}/K_M^+$. It assumes that the rate $v=\partial p/\partial t$ at which the concentration of products $p$ increases depends on the concentration of reactants $s$ and on the total concentration $e_0$ of catalysts by Michaelis-Menten equation~\cite{cornish2014principles},
\beq\label{eq:MMs}
v=\frac{k_{\rm cat}^+e_0s}{K_M^++s}.
\eeq
The ratio $k^+_{\rm cat}/K_M^+$ then characterizes the initial rate of the reaction, when $s\ll K_M^+$. In general, however, Eq.~\eqref{eq:MMs} indicates that the rate $v$ depends on the concentration $s$ of reactants. The ratio $k^+_{\rm cat}/K_M^+$ should indeed be generally interpreted as a measure of specificity rather than a measure of catalytic efficiency~\cite{eisenthal2007catalytic}.

To see how optimizing $k^+_{\rm cat}/K_M^+$ may lead to unphysical results, consider the simplest case where Eq.~\eqref{eq:MMs} arises, under the scheme $E+S  \harp[k_{-1}]{k_{1}} ES\xa{k_2}  E+P$, where the complex $ES$ is assumed to be in a quasi-steady state~\cite{cornish2014principles}. In this case, $k^+_{\rm cat}=k_2$ and $K^+_M=(k_{-1}+k_2)/k_1$. Taking $k_{-1}=0$, we obtain $k^+_{\rm cat}/K^+_M=k_1$, which is independent of $k_2$. Formally, $k^+_{\rm cat}/K^+_M$ can thus be made arbitrarily large by minimizing $k_{-1}$ and maximizing $k_1$, irrespective of $k_2$, even though $k_2$ controls an essential step and $k_2=0$ means that no catalysis takes place. The catch is in the assumption $s\ll K^+_M$, which underlies the choice of the ratio $k^+_{\rm cat}/K^+_M$ as a measure of catalytic efficiency. When $k_{-1}=0$, this assumption implies $s\ll k_2/k_1$, which depends on $k_2$ and is certainly not satisfied when $k_2=0$. This simple example illustrates the need to consider explicitly the concentration $s$ of reactants to obtain physically meaningful results~\footnote{One could also ignore $K_M^+$ and score catalytic efficiency by $k_{\rm cat}^+$ but this choice would not account for the rate at which the product is generated.}. As a corollary a family of optimal designs is defined, which depend on the concentration $s$ of reactants, and not just on the mechanisms of the reaction. More generally, optimal designs also depend on the concentration $p$ of products, which is assumed to be  $p=0$ in Eq.~\eqref{eq:MMs}.

Here, we choose to treat the concentrations and of reactants $s$ and products $p$ as two fixed parameters and to score catalytic activity by the rate $v=\partial p/\partial t$ at which the product is formed. This assumes a reservoir of reactants and products, so that their concentrations are constant despite the reactions that consume or produce them. This is, however, not the only possible choice. One may alternatively consider a closed-system with an initial concentration of reactants and score the concentration of products after a fixed time, or consider a chemostat with a fixed in-flow of reactants and catalysts, a fixed dilution rate and score the out-flow of products.

\section{Solvable one-dimensional model}

The model presented in Figure~\ref{fig:scheme} is defined in any dimension. We study it here in one dimension, where it has only three independent internal degrees of freedom and can be solved analytically. The details of this solution are presented in the appendices and we focus below on the results and assumptions on which they rely. While these assumptions constrain the range of examined designs, they are justified {\it a posteriori} by the finding of  locally optimal designs within their range of validity.

\subsection{Uncatalyzed reaction}

In one dimension, a substrate is characterized by a single internal degree of freedom, the distance $x_0$ between its two beads, and five physical parameters, the spring constants $k_a$ and $k_r$ of the two springs that connect the two beads, their free lengths $l_a$ and $l_r$, and the maximal extension $z_a$ of the breakable spring ($a$ stands for ``attractive'' and $r$ for ``repulsive''). Without loss of generality, we assume $l_a=0$ (Table~\ref{tab:para}). The number of parameters can be further reduced to two by considering adimensional quantities (Appendix~\ref{app:spon}).

\begin{figure}[t]
\begin{center}
\includegraphics[width=.65\linewidth]{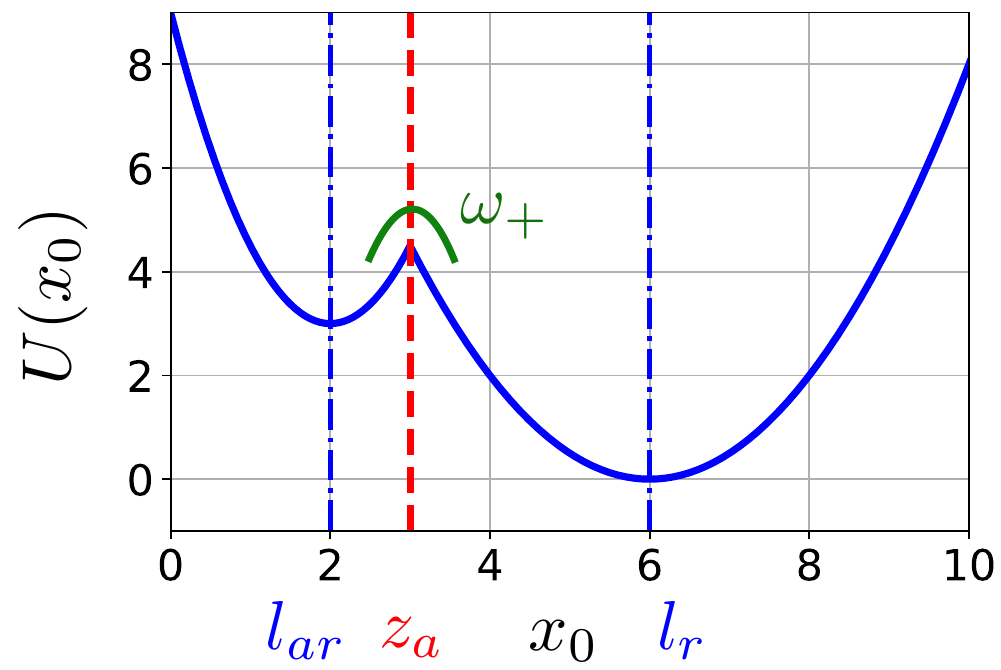}
\caption{Potential for the uncatalyzed reaction $S\harp[]{}P$ -- The potential $U(x_0)$ is a function of the extension $x_0$ of the substrate. The two states $S$ and $P$ are defined by $x_0<z_a$ and $x_0>z_a$ respectively, with the transition between the two defining the reaction $S\harp[]{}P$. The parameters (Table~\ref{tab:para}) for this graph are $k_a=2$, $z_a=3$, $l_r=6$, $k_r=1$. When computing escape rates, we assume a smooth curvature $\omega_+$ at the transition state $x_0=z_a$, where the value of $\omega_+$ is fixed independently of the other parameters (Appendix~\ref{app:kramers}). \label{fig:spont}}
\end{center} 
\end{figure}

As long as the distance $x_0$ between the two beads satisfies $|x_0|<z_a$, the two springs are present and equivalent to a single spring with effective parameters
\beq\label{eq:lar}
k_{ar}=k_a+k_r,\qquad l_{ar}=\frac{k_rl_r}{k_a+k_r}.
\eeq
We assume $0<l_{ar}<z_a< 2l_{ar}$
so that a substrate with initial extension $x_0=l_{ar}$ is more likely to break ($x_0>z_a$) than to invert the relative position of its two beads ($x_0<0$); in this approximation, the interaction potential between the beads is harmonic (Appendix~\ref{app:spon}). For the reactant and the product to be stable, the equilibrium distance with and without the scissile bond must be respectively below and beyond the breaking point, which imposes $l_{ar}<z_a<l_r$. Additionally, we choose parameters so that the state with a broken bond is the state of lowest energy (Appendix~\ref{app:spon} and Fig.~\ref{fig:spont}).  

We compute the rates of transition between states using Kramers' escape formula~\cite{kramers1940brownian}, which assumes that the time scales of relaxation within each state are much smaller then the transition rates. This is valid provided barrier heights are large compared to $k_BT$, where $T$ is the temperature and $k_B$ Boltzmann's constant (Appendix~\ref{app:kramers}). This leads to the forward and reverse rates $\rho_u^+$ (for $S\to P$) and $\rho_u^-$ (for $P\to S$) given by
\begin{align}
\rho_u^+&=\sqrt{k_{ar}}e^{\beta k_{ar}(z_a-l_{ar})^2/2},\nonumber\\ 
\rho_u^-&=\sqrt{k_{r}}e^{\beta k_r (z_a-l_{r})^2/2},
\end{align}
where $\beta=(k_BT)^{-1}$. In these formulae, the unit of time is chosen so that the viscosity $\gamma$ of the solvent and the curvature $\omega_+$ of the potential at the barrier do not appear explicitly (Appendix~\ref{app:kramers}). Given these rates, the reaction $S\to P$ is thermodynamically favored provided $p/s<K_{\rm eq}$ where $s$ and $p$ are the concentrations of the reactant $S$ and product $P$, and where $K_{\rm eq}=\rho_u^+/\rho_u^-$ is the equilibrium constant of the reaction. 

In what follows, we consider as parameters of the reaction (Table~\ref{tab:para}) the values $k_a=2$, $z_a=3$, $l_r=6$, $k_r=1$ and $\beta=2$ so that $l_{ar}=2$ and $k_{ar}=3$ in Eq.~\eqref{eq:lar}. These values, which satisfy the different assumptions that we make (Fig.~S\ref{fig:spcond}), correspond to the potential shown in Figure~\ref{fig:spont}.

\subsection{Catalysis}

The catalyst is characterized by the spring constant $k_e$ and free length $l_e$ of the unbreakable spring that connects its two beads (Fig.~\ref{fig:scheme}). Each of these beads can interact with only one bead of the substrate and the two interactions are described by equivalent breakable springs with spring constant $k_i$, free length $l_i=0$ and maximal extension $z_i$ (Table~\ref{tab:para}). We assume that the catalyst is rigid enough to maintain the relative position of its beads ($k_BT\ll k_el_e^2/2$, Appendix~\ref{app:int}).

The system formed by the catalyst and the substrate can possibly be in $2^3$ states, depending on whether each of the 3 scissile bonds is broken or not. Given the equivalence between the two bonds by which the substrate and the catalyst interact, these 8 states define 6 physically distinct states (Appendix~\ref{app:states} and Fig.~\ref{fig:scheme}). These physical states are well-defined if they are associated with local minima of the potential, and we consider parameters for which this is the case (Appendix~\ref{app:states}).

When all 6 states are well-defined, the catalysis is the result of the series of reactions
\beq\label{eq:chain}
E+S  \harp[\r_{1}^-]{\r_0^+} ES' \harp[\r_{2}^-]{\r_1^+}  ES \harp[\r_{3}^-]{\r_2^+}  EP \harp[\r_{4}^-]{\r_3^+}  EP'\harp[\r_{5}^-]{\r_4^+}  E+P
\eeq
where the intermediate states $ES'$, $ES$, $EP$, $EP'$ are illustrated in Figure~\ref{fig:scheme}.
The transitions $ES' \harp[]{}EP'$ are ignored, which is justified when the rates $\r_u^\pm$ of the uncatalyzed reaction are negligible compared to the rates of the catalyzed reaction, i.e., $\r^+_1\gg\r_u^+$ and $\r^-_4\gg\r_u^-$. We assume $\r_0^+=1$ and $\r_5^-=1$~\footnote{We can always redefine the concentrations $s$ and $p$ so that it is the case. When optimizing at given values of $s$ and $p$, however, this rescaling matters. A non-equivalent choice would for instance be to take $\r_0^+=\r_5^-=4z_i$, with $4z_i$ representing the ``cross-section'' for the collision between catalysts and substrates.} and obtain the other rates by application of Kramers' escape formula (Appendix~\ref{app:transitions}).

Under the assumptions that the concentrations $e_0$ of catalysts (under their different forms), $s$ of reactants and $p$ of products are maintained constant and that the concentrations of all intermediates are at steady state, the rate $v=\partial p/\partial t$ of product formation takes the form
(Appendix~\ref{app:reacrate})
\beq\label{eq:MM}
\frac{v}{e_0}=\frac{k_{\rm cat}^+s/K_M^+-k_{\rm cat}^-p/K_M^-}{1+s/K_M^++p/K_M^-}.
\eeq
The parameters of this reversible Michaelis-Menten equation~\cite{cornish2014principles} depend on the 8 spring parameters given in Table~\ref{tab:para} via the rates in Eq.~\eqref{eq:chain}. They also depend on the temperature of the solvent but not on its viscosity, nor on the curvature of the potential near the activation barriers, which we assume to be identical for all barriers (Appendix~\ref{app:kramers}). 

\section{Optimal designs of 1D catalysts}

\begin{figure}[t]
\begin{center}
\includegraphics[width=\linewidth]{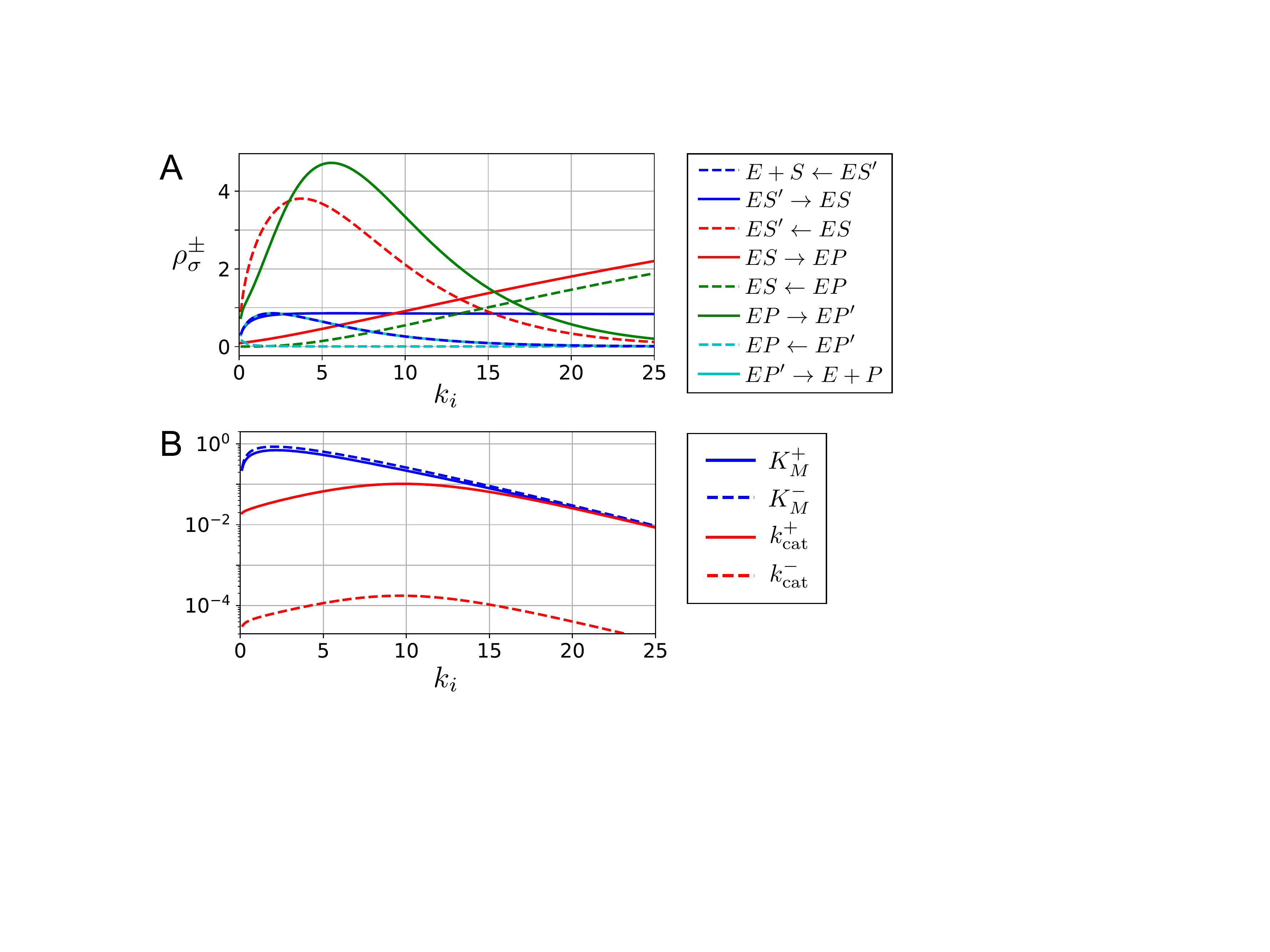}
\caption{Rates of transition between states as a function of the flexibility $k_i$ of the interaction between substrates and catalysts -- {\bf A.}~Rates $\rho_\sigma^\pm$ for the series of transitions given in Eq.~\eqref{eq:chain}. As $k_i$ increases, the rate $\rho_2^+$ of forward catalysis $ES\to EP$ increases (full red line), along with the rate $\rho_3^-$ of reverse catalysis $ES\leftarrow EP$ (dotted green line), but the rates of product release $EP\to EP'$ (full green line) and $EP'\to E+P$ (full cyan line behind the blue dotted line) decrease. {\bf B.}~Michaelis-Menten parameters defined by Eq.~\eqref{eq:MM}. Each parameter has a maximum for an intermediate value of $k_i$. In these graphs, the parameters of the substrate are as in Fig.~\ref{fig:spont} and those of the catalyst other than $k_i$ are given by Eq.~\eqref{eq:tt}. Note that the rates are not independent but satisfy $\prod_\sigma \rho_\sigma^+/\prod_\sigma \rho_\sigma^-=(k_{\rm cat}^+K_M^-)/(K_M^+k_{\rm cat}^-)=K_{\rm eq}$ where $K_{\rm eq}$ is the equilibrium constant of the uncatalyzed reaction (Haldane relationship).\label{fig:rates}}
\end{center} 
\end{figure}

To characterize optimal designs within the model, we maximize the reaction rate $v$ over the four parameters of the catalyst: $k_e,l_e$, which characterize its flexibility and geometry, and $k_i,z_i$, which characterize the strength and range of its interaction with the substrate (Table~\ref{tab:para}). The optimum generally depends on the four physical parameters of the substrate, $k_a,z_a,k_r,l_r$  (Table~\ref{tab:para}), on the concentrations $s$, $p$ at which the reactant $S$ and product $P$ are present, and on the temperature $T$ of the solvent, represented by $\beta=1/(k_BT)$.

For the substrate, we consider the parameters of Figure~\ref{fig:spont}, $k_a=2$, $z_a=3$, $k_r=1$, $l_r=6$, which correspond to parameters $k_{ar}=3$ and $l_{ar}=2$ for the effective bond of the reactant [Eq.~\eqref{eq:lar}]. For the medium, we first consider the parameters $s=10^{-1}$, $p=0$ and $\beta=2$. With these values, we find a locally optimal design (Fig.~S\ref{fig:localopt}) that satisfies all the assumptions involved in the derivation of the rate $v$: $\hat k_e=\infty, \hat l_e=3, \hat k_i\simeq 13,\hat z_i=0.5$.

\begin{figure}[t]
\begin{center}
\includegraphics[width=.75\linewidth]{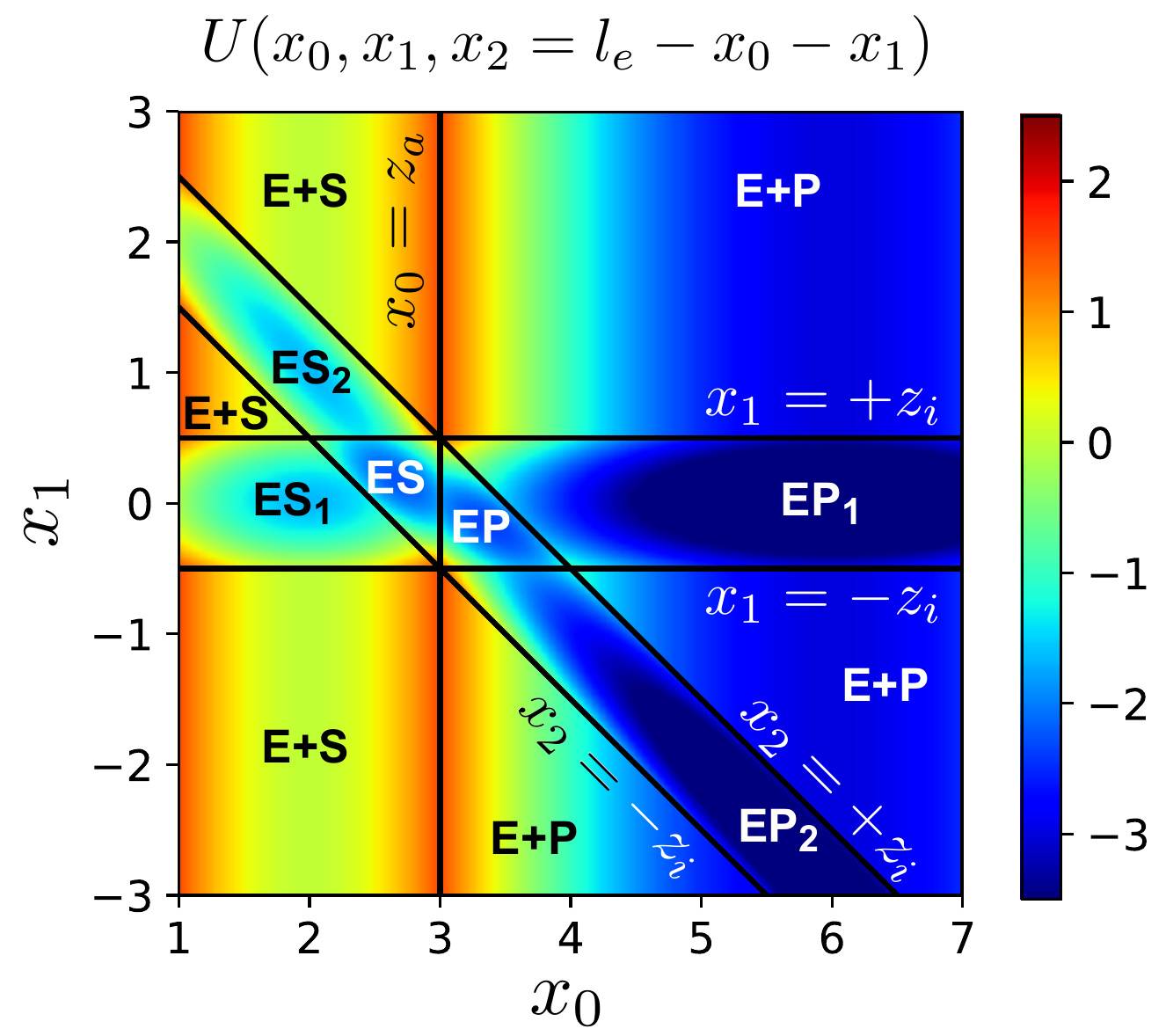}
\caption{Energy landscape of a system substrate-catalyst for an optimal catalyst with infinite rigidity -- The two degrees of freedom are the distance $x_0$ between the two beads of the substrate (the reaction coordinate) and the relative position $x_1$ between a bead of the catalyst and the bead of the substrate with which it interacts. The relative position $x_2$ between the other bead of the catalyst and the other bead of the substrate is given by $x_2=\hat l_e-x_0-x_1$ where $\hat l_e$ is the fixed length of the rigid catalyst. The different states are separated by black lines corresponding to the thresholds beyond which one of the three scissile bonds of the model ruptures: $x_0=z_a$, $x_1=\pm z_i$ and $x_2=\pm z_i$. Here, we distinguish between the two states $ES_1,ES_2$ and $EP_1,EP_2$ instead of subsuming them under common states $ES'$ and $EP'$. The parameters are as in Figure~\ref{fig:rates} with $k_i=13$ and the reference $U=0$ is taken to corresponds to the minimal energy of the state $E+S$.\label{fig:pot2d}}
\end{center} 
\end{figure}

This solution is consistent with the proposal that an optimal catalyst must stabilize the transition state of the reaction~\cite{pauling1948nature,lienhard1973enzymatic}: the catalyst is maximally rigid ($\hat k_e=\infty$) with a length that matches that of the transition state ($\hat l_e=z_a$). Additionally, the range of interaction $\hat z_i$ is adapted to the free length of the substrate: $\hat l_e-2\hat z_i=l_{ar}$. The value of the optimal interaction strength $\hat k_i$ is, on the other hand, less obvious to interpret. It takes a finite value, contrary to what a na\"ive application of the principle of transition-state stabilization would predict. The optimal value of $k_i$ represents indeed a trade-off between the need to stabilize the transition state, which requires rigidity, and the need to release the product, which requires flexibility (Fig.~\ref{fig:rates}). The energy landscape associated with this optimal design can be represented in two dimensions as a rigid catalyst with $\hat k_e=\infty$ leaves only two independent internal degrees of freedom (Fig.~\ref{fig:pot2d}).

Varying the different parameters around the above values, we verify that the relationships associated with transition-state stabilization, 
\beq\label{eq:tt}
\hat k_e=\infty,\qquad \hat l_e=z_a,\qquad \hat z_i=\frac{z_a-l_{ar}}{2}
\eeq
are always nearly satisfied, while $\hat k_i$ is, on the other hand, parameter-dependent (Figs.~S\ref{fig:optext}-S\ref{fig:optphys}). We analyze in what follows the determinants of the optimal interaction strength $\hat k_i$ assuming that the other parameters of the catalyst are given by Eq.~\eqref{eq:tt}.

\subsection{Dependence on concentrations}

\begin{figure}[t]
\begin{center}
\includegraphics[width=\linewidth]{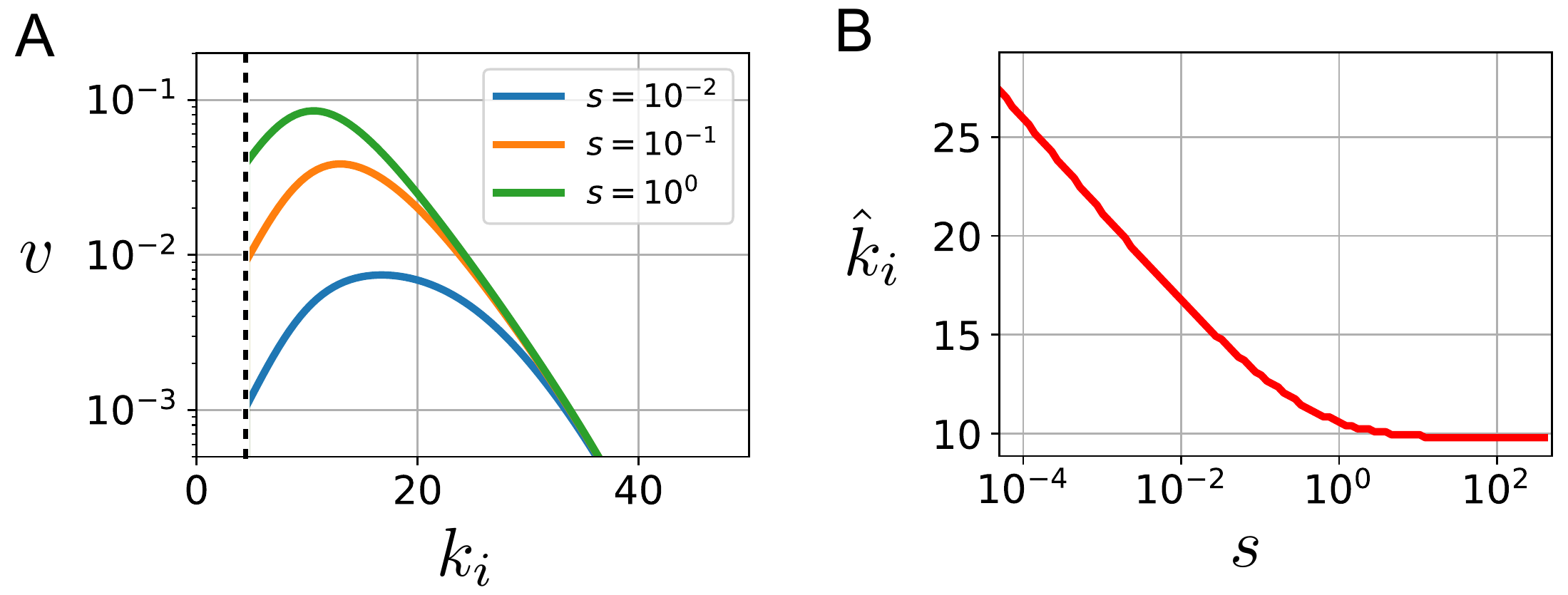}
\caption{{\bf A.} Reaction rate $v$ for the catalyzed reaction as a function of the interaction strength $k_i$ for three different concentrations $s$ of the reactant (and no product, $p=0$), showing that the optimal value of $k_i$ depends on $s$. For $k_i$ smaller than the dashed vertical line, the state $EP$ is unstable and the reaction does not follow the scheme of Eq.~\eqref{eq:chain}.  {\bf B.}~Optimal interaction strength $\hat k_i$ as a function of $s$. \label{fig:ki}}
\end{center} 
\end{figure}

Varying the concentration $s$ of reactants at vanishing concentration of products ($p=0$), we find that $k_i$ has a non-trivial maximum $\hat k_i$ that decreases with $s$ (Fig.~\ref{fig:ki}). In particular, in the limit $s\to 0$ where the problem is equivalent to optimizing the specificity constant $k_{\rm cat}^+/K_M^+$, we have $\hat k_i\to\infty$: the strength of the interaction between substrate and catalyst becomes infinite. This result illustrates how optimizing the ratio $k_{\rm cat}^+/K_M^+$ can lead to unphysical designs as, in this limit, a catalyst is unable to release its product (Fig.~\ref{fig:rates}).

\begin{figure}[t]
\begin{center}
\includegraphics[width=\linewidth]{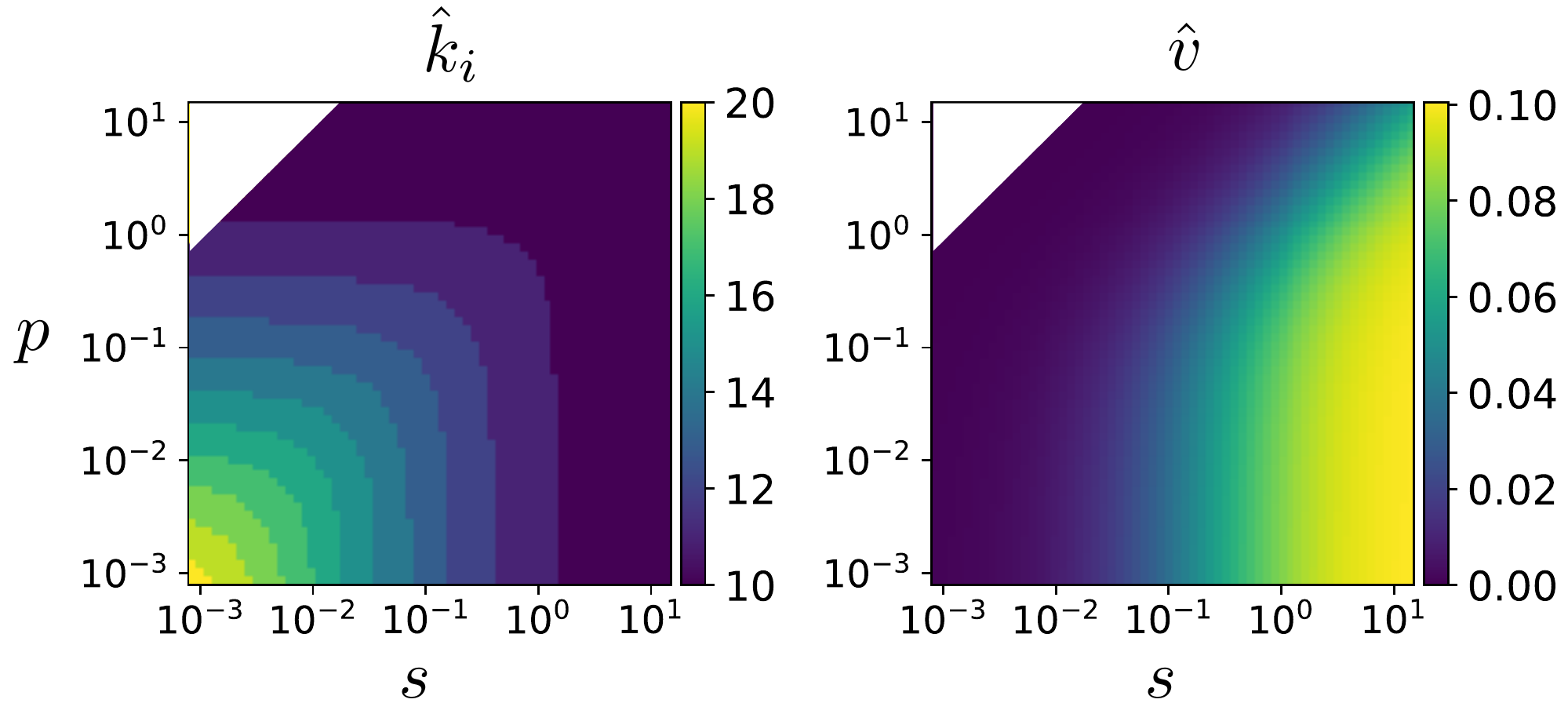}
\caption{Optimal values of the interaction strength $\hat k_i$ and optimal reaction rate $\hat v$ as a function of the concentration $s$ of reactants and $p$ of products. The white triangle in the upper left corner corresponds to $p/s>K_{\rm eq}$, where $K_{\rm eq}$ is the equilibrium constant of the uncatalyzed reaction $S  \harp[]{} P$, in which case the reaction rate $v$ cannot possibly be positive.\label{fig:optsp}}
\end{center} 
\end{figure}

A non-zero concentration of products ($p\neq 0$) introduces an additional constraint, product inhibition. For catalysis to take place, $p$ should be small enough for the reaction to be thermodynamically favored: $p/s<K_{\rm eq}$, where $K_{\rm eq}=\rho_u^+/\rho_{u}^-$ is the equilibrium constant of the uncatalyzed reaction $S  \harp[\r_{u}^-]{\r_u^+} P$ . Under this condition, we find that $\hat k_i$ is a decreasing function of both $s$ and $p$ (Fig.~\ref{fig:optsp}). 

\subsection{Dependence on physical parameters}

\begin{figure}
\begin{center}
\includegraphics[width=.9\linewidth]{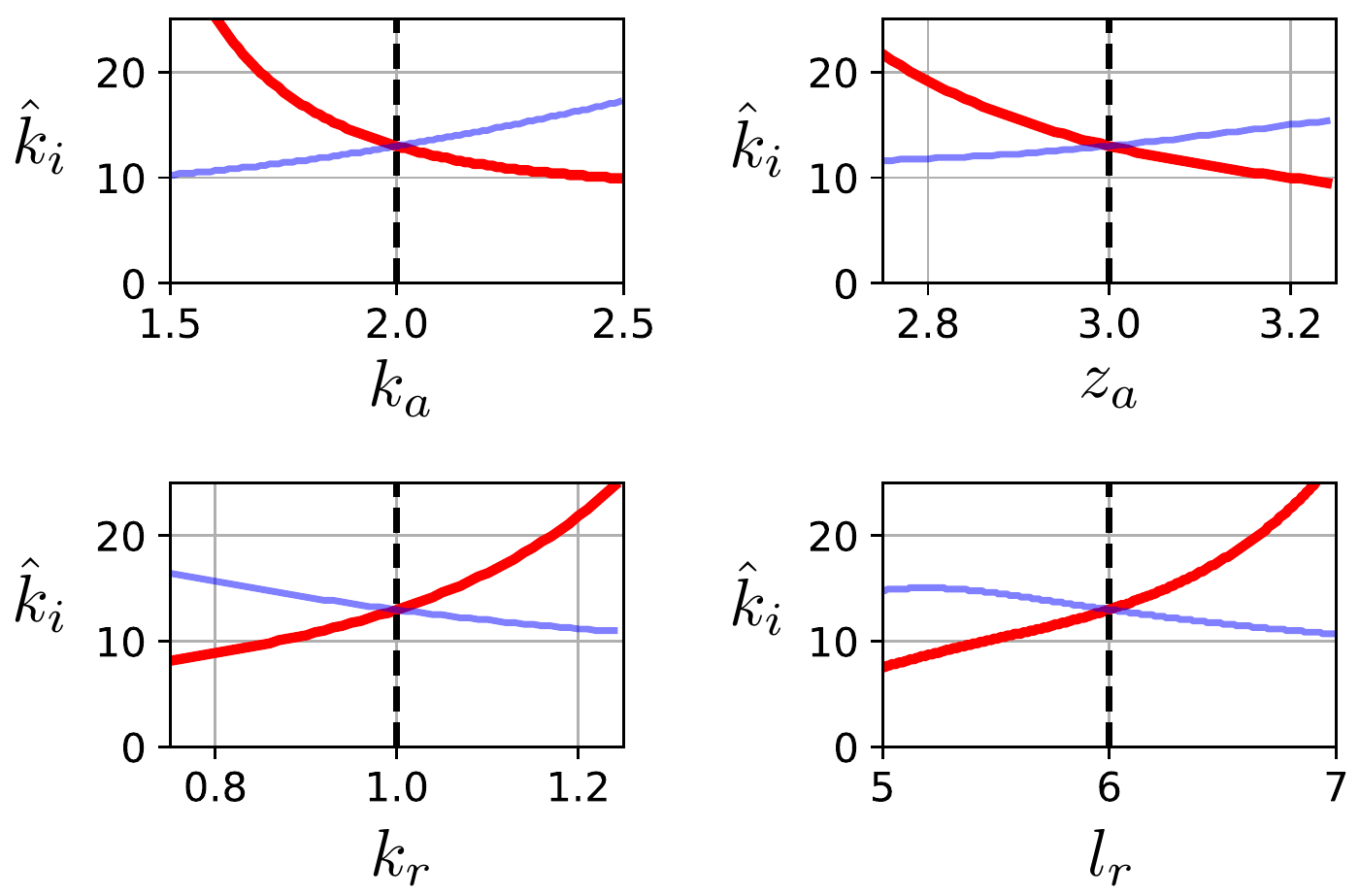}
\caption{Optimal interaction strength $\hat k_i$ (in red) as a function of the physical parameters of the substrate, the strength $k_a$ of the scissile bond, its maximal extension $z_a$, the strength $k_r$ of the non-scissile bond and its extension $l_r$ (Table~\ref{tab:para}). When varying a parameter of the substrate, all the optimal parameters of the catalyst, and not only $\hat k_i$, generally take different values. If fixing these other variables and optimizing over $\hat k_i$ only as in Eq.~\eqref{eq:tilde}, one obtains the blue curves that show opposite trends. The top graphs can also be related to the bottom graphs by noticing that the problem depends on the parameters $k_a,z_a,k_r$ and $l_r$ through the dimensionless quantities $k_a/k_r$ and $l_r/z_a$ [Eq.~\eqref{eq:adim}].
\label{fig:depara}}
\end{center} 
\end{figure}

The dependence of $\hat k_i$ on the physical parameters of the substrate $k_a,z_a,k_r,l_r$ (Table~\ref{tab:para}) is shown in Figure~\ref{fig:depara}. The results are at first sight counter-intuitive. When increasing $k_a$, for instance, the activation  barrier becomes higher but the interaction strength $\hat k_i$ of the optimal catalyst becomes weaker. Similarly, increasing $z_a$ increases the activation barrier but is again associated with a smaller $\hat k_i$. On the other hand, substrates with increased $k_r$ or $l_r$ have a lower activation barrier but are associated with a larger $\hat k_i$.

To rationalize these results, note that varying $k_a,z_a,k_r$ or $l_r$ implies not only a different optimal interaction strength $\hat k_i$ but, from Eq.~\eqref{eq:tt}, a different optimal extension $\hat l_e=z_a$ and a different optimal interaction range $\hat z_i=(z_a-l_{ar})/2$ (Figs.~S\ref{fig:optext}-S\ref{fig:optphys}). If instead of considering
\beq
\hat k_i(k_a)=\arg\max_{k_i} v(z_i=(z_a-l_{ar})/2,k_a)
\eeq
where $l_{ar}$ depends on $k_a$ [Eq.~\eqref{eq:lar}],
as in the red curve of the first panel of Figure~\ref{fig:depara}, we consider 
\beq\label{eq:tilde}
\tilde k_i(k_a)=\arg\max_{k_i} v(z_i,k_a)
\eeq
where $z_i$ is fixed, we obtain the blue curve, which is an increasing function of $k_a$. Mathematically, the observation that stronger bonds are best broken by catalysts making weaker interactions with their substrate is thus explained by the difference between optimizing over a single variable {\it versus} optimizing over all variables jointly. Physically, a stronger $k_a$ reduces the equilibrium length $l_{ar}$ of the reactant and the catalyst needs to be more flexible to bind both to this smaller reactant and to the transition state whose location $z_a$ is unchanged. Reasoning on just one parameter may thus be misleading because varying this parameter may have an incidence on multiple steps of the catalytic cycle and some of these effects may be  compensated by varying other parameters.  {\it Mutatis mutandis}, similar arguments explain the non-trivial dependence on the other parameters shown in Figure~\ref{fig:depara}.

\section{Discussion}

We introduced a simple but general elastic network framework for studying the geometrical and physical constraints to which efficient catalysts are subject and illustrated it with the analytical solution of an elementary one-dimensional model.

The solution demonstrates the relevance and limitations of the principle of transition-state stabilization, which reduces catalysis to binding to (analogues of) the transition state of the reaction~\cite{pauling1948nature,lienhard1973enzymatic}. While we find that the geometry of optimal catalysts matches the geometry of the transition state, consistent with this principle, we also find that binding to this state should not be maximized. Instead, some flexibility is needed to bind to the reactant and release the product in addition to stabilize the transition state. The additional constraints that these requirements impose might explain why catalytic antibodies selected for transition-state stabilization with no consideration of product release are only modest catalysts~\cite{hilvert2000critical}. Binding to the reactant less than to the transition state but more than to the product, which are all chemically similar, poses a problem of fine discrimination. As previously proposed~\cite{rivoire2018minimal}, physical solutions to such problems can rely on a conformational switch: this is the case in the present model where the relative positions of the beads of the catalyst and the substrate are swapped during the transition $ES  \harp[]{} EP$ (Fig.~\ref{fig:scheme}).

While the model is not meant to make quantitative predictions, we note that the optimal strength of interaction between substrate and catalyst is systematically larger than the strength of the bond to break; for instance, in Figure~\ref{fig:rates}B, $k_{\rm cat}^+$ is maximal for $\hat k_i\simeq 5\ k_a$. This is in contrast to enzymes, which can catalyze the rupture of covalent bonds by means of weaker non-covalent interactions. Introducing physical limitations on the strength and length of the various bonds may thus contribute to explain why enzymes are so large~\cite{Srere:1984wf} and why they make multiple interactions with their substrate. This line of reasoning was first followed by Gavish who estimated how much stress an enzyme can exert on a substrate based on a similar toy model~\cite{Gavish86}; his analysis, however, does not consider the full catalytic cycle and, in particular, the need for the catalyst to be flexible to release the product. Besides physical limitations, evolutionary limitations, in particular the granularity of the sequence space, may also be relevant to these questions~\cite{rivoire2018minimal}.

Our model captures another feature of catalysis that is likely to be very general: efficient catalysts are not only optimized for the reaction but for the conditions under which catalysis occurs. In the model, these conditions include the temperature and the concentrations of reactants and products, on which the optimal degree of flexibility $\hat k_i$ depends. In another set-up, these concentrations may not be maintained constant and other parameters may be relevant, such as the concentration of catalysts or the fluctuations due to low concentrations of reactants~\cite{barato2015universal}. 

At a physical level, approximating a molecule by an elastic network is obviously an extreme oversimplification. Enzymes, in particular, are arguably not purely mechanical devices but as importantly electronic devices. Harmonic potentials may describe small distortions of charge distributions as well as mechanical strain, but their particular form, as our simple treatment of the solvent~\cite{Min:2008jb} or our omission of quantum effects~\cite{kohen1998enzyme} certainly limit us to a subset of possible designs.

Within our mechanical framework, several extensions of the model may, however, already be of interest. First, our solution applies only under a number of assumptions that guarantee a sequence of transitions, each described by Kramers' theory~\cite{kramers1940brownian}. We showed that a locally optimal solution exists within the range of validity of these assumptions but did not exclude other solutions beyond this range. Several additional constraints that are relevant to enzymes would also be interesting to incorporate, such as constraints on specificity for the substrate~\cite{fersht1999structure} or long-term evolutionary constraints~\cite{hemery2015evolution}. But going beyond one dimension is maybe the most obvious next step, as a mechanical catalyst must not only apply sufficient strain but orient this strain, which is trivial in one dimension but not in two or three dimensions. Extensions of our model may thus provide further insights on the physical principles of catalysis.

\acknowledgments
This work benefited from stimulating discussions with Cl\'ement Nizak and Zorana Zeravcic and from comments by Eric Rouviere.

\newpage

\clearpage
\newpage
\onecolumngrid
\appendix

\makeatletter
\makeatletter \renewcommand{\fnum@figure}
{\figurename~S\thefigure}
\makeatother
\setcounter{figure}{0}

\renewcommand\theequation{S\arabic{equation}}
\setcounter{equation}{0}

\begin{center}
{\bf APPENDICES}
\end{center}

\subsection{Uncatalyzed reaction}\label{app:spon}

In one dimension, the conformation of the substrate is characterized by the positions of its two beads $x_{s1}$ and $x_{s2}$. The relevant degree of freedom is the distance between them, $x_0=x_{s2}-x_{s1}$. When $|x_0|<z_a$, the two springs are present and the potential is of the form
\beq\label{eq:U0}
U(x_0)=\frac{k_{ar}}{2}(|x_0|-l_{ar})^2+C_{ar}
\eeq
where
\beq
k_{ar}=k_a+k_r,\qquad l_{ar}=\frac{k_rl_r}{k_a+k_r},
\eeq
and where $C_{ar}$ is an arbitrary constant. We assume
\beq
0<l_{ar}<z_a< 2l_{ar}
\eeq
so that $U(z_a)<U(0)$ and a substrate with initial extension $x_0=l_{ar}$ is more likely to break ($x_0>z_a$) than to invert the relative position of its two beads ($x_0<0$). Under this assumption, Eq.~\eqref{eq:U0} can be simplified to the harmonic potential
\beq
U(x_0)=\frac{k_{ar}}{2}(x_0-l_{ar})^2+C_{ar}.
\eeq

When $x_0>z_a$, only one spring is present and the potential becomes
\beq
U(x_0)=\frac{k_{r}}{2}(x_0-l_{r})^2+C_{r}
\eeq
where $C_r$ is related to $C_{ar}$ by a condition of continuity at $x_0=z_a$. 

For the reactant and the product to be stable, the equilibrium points with and without the breakable spring must be respectively below and beyond the breaking point, which imposes $l_{ar}<z_a<l_r$. Additionally, requiring the product to be the state of minimal energy imposes $U(l_r)<U(l_{ar})$, i.e., $k_{ar}(z_a-l_{ar})^2<k_{r}(z_a-l_r)^2$.  

Finally, we assume that the relaxation time is much smaller than the escape time, $U(z_a)-U(l_{ar})\gg k_BT$, i.e., $k_{ar}(z_a-l_{ar})^2/2/\gg \beta^{-1}$. We can then apply Kramers' escape formula (Appendix~\ref{app:kramers}) to obtain the forward ($S\to P$) and reverse ($P\to S$) rates of the uncatalyzed reaction as
\beq
\rho_u^+=\sqrt{k_{ar}}e^{\beta k_{ar}(z_a-l_{ar})^2/2},\qquad \rho_u^-=\sqrt{k_{r}}e^{\beta k_r (z_a-l_{r})^2/2}
\eeq
where $\beta=(k_BT)^{-1}$. The unit of time is chosen here so that the viscosity $\gamma$ of the solvent and the curvature $\omega_+$ of the potential at the barrier do not appear explicitly (Appendix~\ref{app:kramers}).

\begin{figure}[t]
\begin{center}
\includegraphics[width=.8\linewidth]{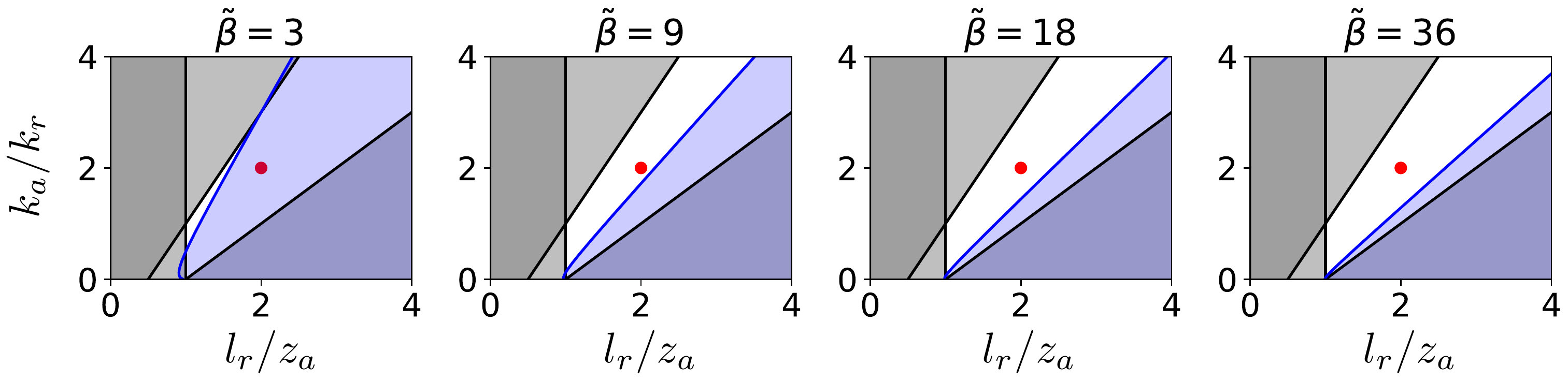}
\caption{For the spontaneous reaction to satisfy the assumptions given in Eq.~\eqref{eq:assumptions}, the parameters $k_a,z_a,k_r,l_r,\beta$ must be chosen within the white region [see Eq.~\eqref{eq:spcd}]. The choice made in the main text corresponds to the red dot in the graph with $\tilde\beta=18$. \label{fig:spcond}}
\end{center} 
\end{figure}

The uncatalyzed reaction involves 5 parameters, $k_a,z_a,k_r,l_r,\beta$, but the 4 different assumptions 
\beq\label{eq:assumptions}
z_a<l_r,\quad l_{ar}<z_a,\quad z_a<2l_{ar},\quad k_{ar}(z_a-l_{ar})^2<k_{r}(z_a-l_r)^2\quad 1<\beta k_{ar}(z_a-l_{ar})^2/2,
\eeq
can be formulated in terms of just 3 adimensional parameters,
\beq\label{eq:adim}
\tilde k=k_a/k_r,\quad\tilde \ell=l_r/z_a,\quad\tilde\beta=\beta k_az_a^2/2,
\eeq
as
\beq\label{eq:spcd}
1<\tilde  \ell,\quad \tilde  \ell-1<\tilde k,\quad 2\tilde k <\tilde  \ell-1,\quad 1-\tilde k^2<\tilde\ell,\quad
\tilde  \ell<(1+\tilde k)\left[1-\left(\tilde\beta(1+\tilde k^{-1})\right)^{-1/2}\right].
\eeq
These conditions are represented graphically in Fig.~S\ref{fig:spcond} for different values of $\tilde\beta$ ($1-\tilde k^2<\tilde\ell$ is implied by $1<\tilde  \ell$ and $\ell-1<\tilde k$). The choice $k_a=2$, $z_a=2$, $k_r=1$, $\beta=2$ made in the main text corresponds to the red point at $\tilde l=2$ and $\tilde k=2$ for $\tilde\beta=18$.

\subsection{Kramers' escape rate formula}\label{app:kramers}

Consider a particle in a potential $U(x)$ subject to friction and to a random force satisfying the fluctuation-dissipation theorem. In the limit of strong friction (over-damped regime), its dynamics is 
described by a Langevin equation of the form
\beq
-\zeta \partial_t x=-\partial_xU(x) +\sqrt{2\zeta/\beta}\ \xi(t)
\eeq
where $\zeta$ is the friction, which is proportional to the viscosity $\gamma$ of the solvent, $\beta=1/(k_B T)$ with $k_B$ the Boltzmann constant and $T$ the temperature, and where $\xi(t)$ is a Gaussian white noise with $\langle\xi(t)\xi(t')\rangle=\delta(t-t')$.

We assume that the particle is initially at a local minimum of the potential located at the origin: $x(t=0)=0$ and $\partial_t x(t=0)=0$. The potential is assumed to be smooth, with a local maximum at $z_0$ and we consider the rate $\rho_0$ at which the particle crosses the barrier at $z_0$ to reach a second minimum at $x_1>z_a$. Assuming the barrier height $\Delta U^+=U(z_0)-U(0)$ to verify $\Delta U^+\gg k_BT$, Kramers' escape formula gives in the high-friction limit ($\zeta\gg \omega^+$)~\cite{kramers1940brownian}
\beq
\r_0=\frac{\omega_0\omega_+}{2\pi\zeta}e^{-\beta\Delta U^+}
\eeq
where $\omega_0^2=\partial_x^2U(x=0)$ is the curvature of the potential at the local minimum $x=0$ and $\omega_+^2=-\partial_x^2U(x=z_0)$ at the local maximum $x=z_0$.

This formula cannot be applied directly to a potential of the form $U(x)=k_0x^2/2$ if $x<z_0$ and $U(x)=k_1(x-\ell)^2/2+C$ if $x>z_0,$ since $\omega_+^2=-\partial_x^2U(x=z_0)$ is not defined. As the discontinuity at the barrier is not physically relevant, it is simpler and as relevant to assume that the barrier is smooth with a curvature $\omega^+$ that is independent of the other parameters $k_0$ and $z_0$. We therefore consider as rate of escape
\beq
\r_0=\frac{\omega_+}{2\pi \zeta}\sqrt{k_0} e^{-\beta k_0 z_0^2/2}.
\eeq
As we assume the same curvature $\omega_+$ for all activation barriers, the prefactor $\omega_+/\zeta$ is common to all the reaction rates and we can effectively ignore it by setting the unit of time such that $\omega_+/\zeta=2\pi$.

\subsection{Interaction substrate-catalyst}\label{app:int}

If $x_{s1}<x_{s2}$ and $x_{e1}<x_{e2}$ are, respectively, the positions of the two beads of the substrate and of the catalyst, it is convenient to consider as variables the extension $x_0$ of the substrate and the distances $x_1$ and $x_2$ between the interacting beads of the substrate and catalyst,
\begin{align}
x_0&=x_{s2}-x_{s1},\nonumber\\ 
x_1&=x_{s1}-x_{e1},\\
x_2&=-x_{s2}+x_{e2},\nonumber\
\end{align}
where the last sign is chosen to have $x_{e2}-x_{e1}=x_0+x_1+x_2$.
With these variables, the potential of the system is of the form
\beq\label{eq:U012}
U(x_{0},x_{1},x_{2})=\frac{k_0}{2}(x_0-l_0)^2+\frac{k_1}{2}x_1^2+\frac{k_2}{2}x_2^2
+\frac{k_e}{2}(x_0+x_1+x_2-l_e)^2+C.
\eeq
To specify the parameters in this formula, a total of $2^3=8$ cases must be distinguished, depending on whether each of the 3 breakable springs is formed or not. The values of $k_0$, $l_0$, $k_1$, $k_2$ in each of these cases are given in Table~\ref{tab:Utot}. The constants $C$ also differ in each case and are set to ensure the continuity of the potential. Note that we make here two harmonic approximations, for the substrate and the catalyst, which are justified provided $z_a\ll 2l_{ar}$ ($x_0>z_a$ is more likely than $x_0<0$) and $k_BT\ll k_el_e^2/2$ (the catalyst is rigid enough to be unlikely to go from $x_{e1}<x_{e2}$ to $x_{e1}>x_{e2}$).

\begin{table}
\begin{center}
\begin{tabular}{|l|c|c|c|c|c|c|c|}
  \hline
 & & & & $k_0$ & $l_0$ & $k_{1}$ & $k_{2}$\\
  \hline
$E+S$ & $x_0<z_a$ & $|x_1|>z_i$ & $|x_2|>z_i$ & $k_{ar}$ & $l_{ar}$ & $0$ & 0  \\
$ES_1$ & $x_0<z_a$ & $|x_1|<z_i$ & $|x_2|>z_i$ & $k_{ar}$ & $l_{ar}$ & $k_i$ & 0\\
$ES_2$ & $x_0<z_a$ & $|x_1|>z_i$ & $|x_2|<z_i$ & $k_{ar}$ & $l_{ar}$ & $0$ & $k_i$\\
$ES$ & $x_0<z_a$ & $|x_1|<z_i$ & $|x_2|<z_i$ & $k_{ar}$ & $l_{ar}$ & $k_i$ & $k_i$\\
$EP$ & $x_0>z_a$ & $|x_1|<z_i$ & $|x_2|<z_i$ & $k_{r}$ & $l_{r}$ & $k_i$ & $k_i$\\
$EP_1$ & $x_0>z_a$ & $|x_1|<z_i$ & $|x_2|>z_i$ & $k_{r}$ & $l_{r}$ & $k_i$ & 0\\
$EP_2$ & $x_0>z_a$ & $|x_1|>z_i$ & $|x_2|<z_i$ & $k_{r}$ & $l_{r}$ & $0$ & $k_i$\\
$E+P$ & $x_0>z_a$ & $|x_1|>z_i$ & $|x_2|>z_i$ & $k_{r}$ & $l_{r}$ & $0$ & 0\\
 \hline
\end{tabular}
\end{center}
\caption{Values of the parameters $k_0$, $l_0$, $k_1$, $k_2$ in the formula of the total potential $U(x)$, Eq.~\eqref{eq:U012}. The 8 different cases are defined by the first four columns where $x_0$ is the extension of the substrate and $x_1$, $x_2$ the two distances between the interacting beads of the substrate and catalyst. For instance, $ES_1$ corresponds to a substrate in state $S$ ($x_0<z_a$) that interacts with the catalyst through their first beads ($|x_1|<z_i$) but not through their second ($|x_2|>z_i$), in which case $k_0=k_{ar}$, $l_0=l_{ar}$, $k_1=k_i$ and $k_2=0$, where $l_{ar}$ and $k_{ar}$ are given by Eq.~\eqref{eq:lar}.\label{tab:Utot}}
\end{table}

\subsection{States of the system}\label{app:states}

Given the equivalence of the two bonds by which the substrate and the catalyst can interact, the 8 different states defined in Table~\ref{tab:Utot} represent only 6 physically distinct states. The two states $ES_1$ and $ES_2$ where the substrate $S$ is attached through a single end to the catalyst $E$ can indeed be described by a single state $ES'$, and the two states $EP_1$ and $EP_2$ where $P$ is attached by a single end to $E$ by $EP'$ (Fig.~\ref{fig:scheme}). These states are physically meaningful if they are associated with local minima of the potential, which corresponds to the conditions given by the last columns of Table~\ref{tab:states}, where $L_n$ is in each case the value of $x_n$ that minimizes the potential. The formulae for $L_n$ are
\begin{align}\label{eq:Ln}
L_0&=\frac{k_{-0}l_e+k_0l_0}{K_0},& K_0&=k_{-0}+k_0,\nonumber\\
L_1&=\frac{k_{-1}(l_e-l_0)}{K_1},& K_1&=k_{-1}+k_1,\\
L_2&=\frac{k_{-2}(l_e-l_0)}{K_2},& K_2&=k_{-2}+k_2,\nonumber
\end{align}
with the values of $k_0,l_0,k_1,k_2$ for each state $\s$ given in Table~\ref{tab:states} and with
\beq
k_{-n}^{-1}=k_e^{-1}+\sum_{m\neq n}k_m^{-1}
\eeq
where the sum is over $m=0,1,2$, excluding $m=n$.
For instance, $k_{-1}=(k_e^{-1}+k_0^{-1}+k_2^{-1})^{-1}=k_ek_0k_2/(k_ek_0+k_ek_2+k_0k_2)$, which is 0 when $k_0=k_2=0$. 

\begin{table}
\begin{center}
\begin{tabular}{|l|c|c|c|c|c|c|c|c|}
  \hline
& $\s$ & $k_0$ & $l_0$ & $k_{1}$ & $k_{2}$  & $L_0<z_a$ & $|L_1|<z_i$ & $|L_2|<z_i$ \\
  \hline
$E+S$ & 0 & $k_{ar}$ & $l_{ar}$ & $0$ & 0  &  1 & 0 & 0\\
$ES'$ & 1 & $k_{ar}$ & $l_{ar}$ & $k_i$ & 0  &  1 & 1 & 0 \\
$ES$ & 2 & $k_{ar}$ & $l_{ar}$ & $k_i$ & $k_i$ &  1 & 1 & 1 \\
$EP$ & 3 &$k_r$ & $l_r$ & $k_i$ & $k_i$  &  0 & 1 & 1 \\
$EP'$ & 4 &$k_r$ & $l_r$ & $k_i$ & 0 &  0 & 1 & 0\\
$E+P$ & 5 &$k_r$ & $l_r$ & $0$ & 0 &  0 & 0 & 0 \\
  \hline
\end{tabular}
\end{center}
\caption{Properties of the 6 distinct states $\s$ that the substrate-catalyst complex may take. The first columns give the values that $k_0,l_0,k_1,k_2$  take in the formula for the potential $U(x_0,x_1,x_2)$, Eq.~\eqref{eq:U012}. $ES'$ and $EP'$ each represent two cases, respectively $ES_1,ES_2$ and $EP_1,EP_2$. The three last columns give conditions for the states to be (meta)stable with 0/1 indicating that the condition $L_0<z_a$, $|L_1|<z_i$ or $|L_2|<z_i$ must be violated/satisfied, where the $L_n$ are given by Eq.~\eqref{eq:Ln}.\label{tab:states}}
\end{table}

To obtain these formulae, consider
\beq
V_n(x_n)=\min_{\{x_m\}_{m\neq n}}U(x_0,x_1,x_2)
\eeq
for $U(x_0,x_1,x_2)$ given by Eq.~\eqref{eq:U012}. $V_n(x_n)$ can be rewritten as
\beq
V_n(x_n)=\frac{1}{2}K_n(x_n-L_n)^2+C_n.
\eeq
We derive the expression for $K_n$ and $L_n$ as a function of $k_0,l_0,k_1,k_2,k_e,l_e$ for $n=0,1$ (the case $n=2$ is obtained from the case $n=1$ by exchanging $k_1$ and $k_2$) by repeated use of the formula
\beq\label{eq:addU}
\frac{1}{2}k_1(x-l_1)^2+\frac{1}{2}k_2(x-l_2)^2=\frac{1}{2}k_{1|2}(x-l_{1|2})^2+\frac{1}{2}k_{1-2}l_{1-2}^2
\eeq
with the following notations:
\beq
k_{a|b}=k_a+k_b,\qquad l_{a|b} =\frac{k_al_a+k_bl_b}{k_a+k_b},\qquad k_{a-b}=\frac{k_ak_b}{k_a+k_b},\qquad l_{a-b}=l_b-l_a.
\eeq
This formula implies
\beq\label{eq:minU}
\min_x\left[\frac{1}{2}k_1(x-l_1)^2+\frac{1}{2}k_2(x-l_2)^2\right]=\frac{1}{2}k_{1-2}l_{1-2}^2.
\eeq

We first apply Eq.~\eqref{eq:minU} to obtain
\beq
\min_{x_2}U(x_{0},x_{1},x_{2})=\frac{1}{2}k_0(x_0-l_0)^2+\frac{1}{2}k_1x_1^2+\frac{1}{2}k_{2-e}(x_0+x_1-l_e)^2,
\eeq
and then Eq.~\eqref{eq:addU} followed by Eq.~\eqref{eq:minU} to obtain
\bea
V_0(x_0)=\min_{x_1,x_2}U(x_0,x_1,x_2)&=&\frac{1}{2}k_0(x_0-l_0)^2+\frac{1}{2}k_{1-2-e}(x_0-l_e)^2\\
&=&\frac{1}{2}k_{0|(1-2-e)}\left(x_0-\frac{k_0l_0+k_{1-2-e}l_e}{k_0+k_{1-2-e}}\right)^2+\frac{1}{2}k_{0-1-2-e}(l_e-l_0)^2,\\
V_1(x_1)=\min_{x_0,x_2}U(x_0,x_1,x_2)&=&\frac{1}{2}k_{0-2-e}(x_1-l_e+l_0)^2+\frac{1}{2}k_{1}x_1^2\\
&=&\frac{1}{2}k_{1|(0-2-e)}\left(x_1-\frac{k_{0-2-e}(l_e-l_0)}{k_{1}+k_{0-2-e}}\right)^2+\frac{1}{2}k_{0-1-2-e}(l_e-l_0)^2.
\eea
From these equations, we read
\bea
K_0&=&k_{0|(1-2-e)}=k_0+k_{1-2-e},\qquad L_0=\frac{k_0l_0+k_{1-2-e}l_e}{k_0+k_{1-2-e}},\\
K_1&=&k_{1|(0-2-e)}=k_1+k_{0-2-e},\qquad L_1=\frac{k_{0-2-e}(l_e-l_0)}{k_{1}+k_{0-2-e}}.
\eea
Introducing 
\beq
k_{-n}=\left(k_e^{-1}+\sum_{m\neq n}k_m^{-1}\right)^{-1},
\eeq
we have $k_{1-2-e}=k_{-0}$, $k_{0-2-e}=k_{-1}$, and therefore Eq.~\eqref{eq:Ln}.

\subsection{Transition rates}\label{app:transitions}

Under the assumption that all 6 intermediate states are local minima of the potential, we have the chain of reactions illustrated in Figure~\ref{fig:scheme},
\beq
E+S  \harp[\r_{1}^-]{\r_0^+} ES' \harp[\r_{2}^-]{\r_1^+}  ES \harp[\r_{3}^-]{\r_2^+}  EP \harp[\r_{4}^-]{\r_3^+}  EP'\harp[\r_{5}^-]{\r_4^+}  E+P.
\eeq
Here the transition rates $\rho_\s^\epsilon$ are labeled by the initial state $\s$ and by the direction $\epsilon$ of the transition. Note that the transitions $ES' \harp[]{}EP'$ are ignored, which is justified when the spontaneous rates $\r_0^\pm$ are negligible compared to the catalyzed rates, i.e., $\r^+_1\gg\r_0^+$ and $\r^-_4\gg\r_0^-$.

\begin{table}
\begin{center}
\begin{tabular}{|l|c|c|c|c|}
  \hline
& $\s$ & $\epsilon$ & $n$ & $\mu$\\
  \hline
 $ES'\to E+S$ & 1& -  & 1 & 0\\
 $ES'\to ES$ & 1& + & 2 & 2\\
 $ES\to ES'$ & 2&  - & 2 & 1\\
 $ES\to EP$ & 2&  + & 0 & 3\\
 $EP\to ES$ & 3&  - & 0 & 2\\
 $EP\to EP'$ & 3&  + & 2 & 4\\
 $EP'\to EP$ & 4& - & 2 & 3\\
 $EP'\to E+P$ & 4 &  + & 1 & 5\\
  \hline
\end{tabular}\\
\end{center}
\caption{The 8 possible rates $\rho^\epsilon_\epsilon$ in Eq.~\eqref{eq:chain} are indexed by the initial state $\s$ and a direction $\epsilon=\pm$. They are associated in Eq.~\eqref{eq:rates} with two parameters $n$ and $\mu$ defined in this table. \label{tab:reac}}
\end{table}

We treat each transition as a unidimensional problem of barrier crossing by considering only the most likely trajectory. If we assume that the first beads ($n=1$) are always the first and last to be attached, the relevant dimensions, or ``reaction coordinates'', are $x_0$ ($n=0$) for $ES \harp[\r_{3}^-]{\r_2^+}  EP$, $x_1$ ($n=1$) for $E+S  \harp[\r_{1}^-]{\r_0^+} ES'$, $EP'\harp[\r_{5}^-]{\r_4^+}  E+P$, and $x_2$ ($n=2$) for $ES' \harp[\r_{2}^-]{\r_1^+}  ES$, $EP \harp[\r_{4}^-]{\r_3^+}  EP'$. The effective potential along one of these dimensions is
\beq
V_n(x_n)=\min_{\{x_m\}_{m\neq n}}U(x_{0},x_{1},x_{2})
\eeq
which is equivalently written
\beq
V_{n}(x_n)=\frac{1}{2}K_{n}(x_n-L_{n})^2+C_{n}
\eeq
where $K_{n}$ and $L_{n}$ are given by Eq.~\eqref{eq:Ln}. 

Using Kramers' escape formula (Appendix~\ref{app:kramers}), the rates are then given by
\beq\label{eq:rates}
\rho_\s^\epsilon= q_\mu\sqrt{K_{n}}\exp\left(-\frac{\beta K_{n}(z_{n}-\phi_n(L_{n}))^2}{2}\right).
\eeq
Here, $q_\mu$ is a multiplicity factor that accounts for the fact that $ES'$ and $EP'$ cover two cases; the mapping $(\s,\epsilon)\mapsto \mu$ is given in Table~\ref{tab:reac} and the mapping $\mu\mapsto q_\mu$ in Table~\ref{tab:states}. $n$ labels the reaction coordinate; the mapping $(\s,\epsilon)\mapsto n$ is given in Table~\ref{tab:reac} and the values of $k_0,l_0,k_1,k_2$ to be used in Eq.~\eqref{eq:Ln} to obtain $L_{n}$ and $K_{n}$ are given in Table~\ref{tab:states}. Finally, the mappings $n\mapsto z_n$ and $n\mapsto \phi_n$ are given in Table~\ref{tab:z}; $z_n$ specifies the relevant threshold, $z_a$ or $z_i$ while $\phi_n(x)=x$ or $|x|$. The need for an absolute value arises because two thresholds $\pm z_i$ are involved when considering $n=1,2$; the most relevant is the one that minimizes $V_n(x_n)$, which has the sign of $L_n$, and the formula rests on the observation that $({\rm sign}(L_n)z_n-L_n)^2=(z_n-|L_n|)^2$.

\subsection{Reaction rate}\label{app:reacrate}

The concentrations of the different chemical species satisfy
\bea
\partial_t[S]&=&-\r_0^+[E][S]+\r_1^-[ES'],\nonumber\\
\partial_t[ES']&=&\r_0^+[E][S]-(\r_1^-+\r_1^+)[ES']+\r_2^-[ES]\nonumber,\\
\partial_t[ES]&=&\r_1^+[ES']-(\r_2^-+\r_2^+)[ES]+\r_3^-[EP]\nonumber,\\
\partial_t[EP]&=&\r_2^+[ES]-(\r_3^-+\r_3^+)[EP]+\r_4^-[EP'],\\
\partial_t[EP']&=&\r_3^+[EP]-(\r_4^-+\r_4^+)[EP']+\r_5^-[E][P]\nonumber,\\
\partial_t[P]&=&\r_4^+[EP']-\r_5^-[E][P]\nonumber  .
\eea
By redefining $s=[S]$ and $p=[P]$, we can always assume that $\r_0^+=1$ and $\r_5^-=1$. 

We consider the steady-state solution upon a fixed total concentration of catalysts,
\beq
[E]+[ES']+[ES]+[EP]+[EP']=e_0.
\eeq
The formula for the rate of production $v=\partial_t[P]$ as a function $[S]=s$, $[P]=p$ and $e_0$ can for instance be obtained by the diagrammatic method of King and Altman~\cite{cornish2014principles}. It yields 
\beq\label{eq:v}
\frac{v}{e_0}=\frac{\r_1^+\r_2^+\r_3^+\r_4^+s-\r_1^-\r_2^-\r_3^-\r_4^-p}{\kappa_0+\kappa_ss+\kappa_pp}
\eeq
with
\bea\label{eq:kappa}
\kappa_0&=&\r_1^-\r_2^-\r_3^-\r_4^-+\r_1^-\r_2^-\r_3^-\r_4^++\r_1^-\r_2^-\r_3^+\r_4^++\r_1^-\r_2^+\r_3^+\r_4^++\r_1^+\r_2^+\r_3^+\r_4^+,\\
\kappa_s&=&\r_1^+\r_2^+\r_3^++\r_2^-\r_3^-\r_4^-+\r_2^-\r_3^-\r_4^++\r_2^-\r_3^+\r_4^++\r_2^+\r_3^+\r_4^++\r_1^+\r_3^-\r_4^-+\r_1^+\r_3^-\r_4^++\r_1^+\r_3^+\r_4^++\r_1^+\r_2^+\r_4^-+\r_1^+\r_2^+\r_4^+,\nonumber\\
\kappa_p&=&\r_1^-\r_2^-\r_3^-+\r_1^-\r_2^-\r_3^++\r_1^-\r_2^+\r_3^++\r_1^+\r_2^+\r_3^++\r_2^-\r_3^-\r_4^-+\r_1^-\r_3^-\r_4^-+\r_1^+\r_3^-\r_4^-+\r_1^-\r_2^-\r_4^-+\r_1^-\r_2^+\r_4^-+\r_1^+\r_2^+\r_4^-.\nonumber
\eea
This equation has the form a reversible Michaelis-Menten equation
\beq
\frac{v}{e_0}=\frac{k_{\rm cat}^+s/K_M^+-k_{\rm cat}^-p/K_M^-}{1+s/K_M^++p/K_M^-}
\eeq
where $k_{\rm cat}^+$, $K_M^+$, $k_{\rm cat}^-$ and $K_M^-$ are given by
\beq
k_{\rm cat}^+=\frac{\r_1^+\r_2^+\r_3^+\r_4^+}{\kappa_s},\quad 
k_{\rm cat}^-=\frac{\r_1^-\r_2^-\r_3^-\r_4^-}{\kappa_p},\quad
K_M^+=\frac{\kappa_0}{\kappa_s},\quad K_M^-=\frac{\kappa_0}{\kappa_p}.
\eeq

\begin{table}
\begin{center}
\begin{tabular}{|c|c|c|c|}
  \hline
$n$ & $z_n$ & $\phi_n(x)$\\
  \hline
 0 & $z_a$ & $x$\\
 1 & $z_i$ & $|x|$\\
 2 & $z_i$ & $|x|$\\
  \hline
\end{tabular}\\
\end{center}
\caption{Definitions of $z_n$ and $\phi_n(x)$ used in Eq.~\eqref{eq:rates}, where the mapping $(\s,\epsilon)\mapsto n$ is given in Table~\ref{tab:reac}.\label{tab:z}}
\end{table}

\newpage

\begin{center}
{\bf SUPPLEMENTARY FIGURES}
\end{center}

\begin{figure}[h]
\begin{center}
\includegraphics[width=.85\linewidth]{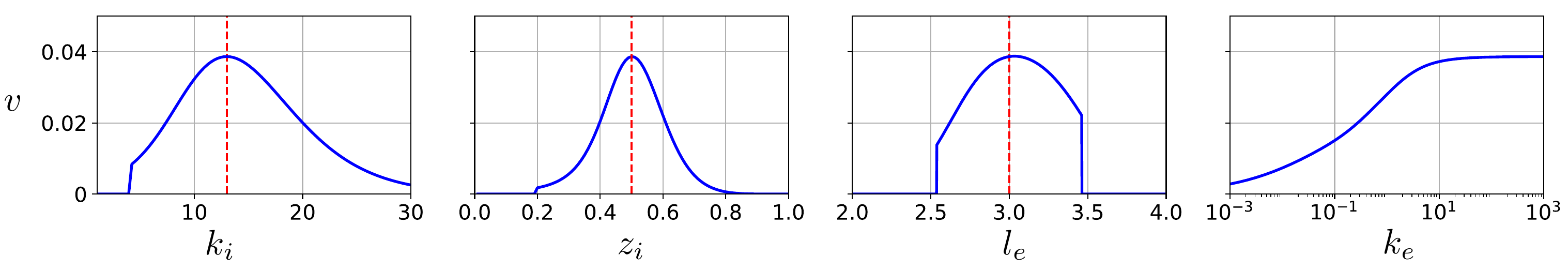}
\caption{Dependence of the rate $v$ of product formation on the different parameters of the catalyst, $k_i$, $z_i$, $l_e$ and $k_e$ for $k_a=2$, $z_a=3$, $k_r=1$, $l_r=6$, $\beta=2$, $s=0.1$, $p=0$ and $e_0=1$. 
Here we consider $\hat z_i=(z_a-l_{ar})/2=3$, $\hat l_e=z_a=3$, $\hat k_e=\infty$ and $\hat k_i\simeq 13$ and vary alternatively each of these parameters. The graphs show that these values of the parameters define a local optimum of $v$. In these graphs, the reaction rate is arbitrarily set to $v=0$ when one of the states becomes unstable and Eq.~\eqref{eq:v} is therefore no longer applicable.\label{fig:localopt}}
\end{center} 
\end{figure}

\begin{figure}[h]
\begin{center}
\includegraphics[width=.45\linewidth]{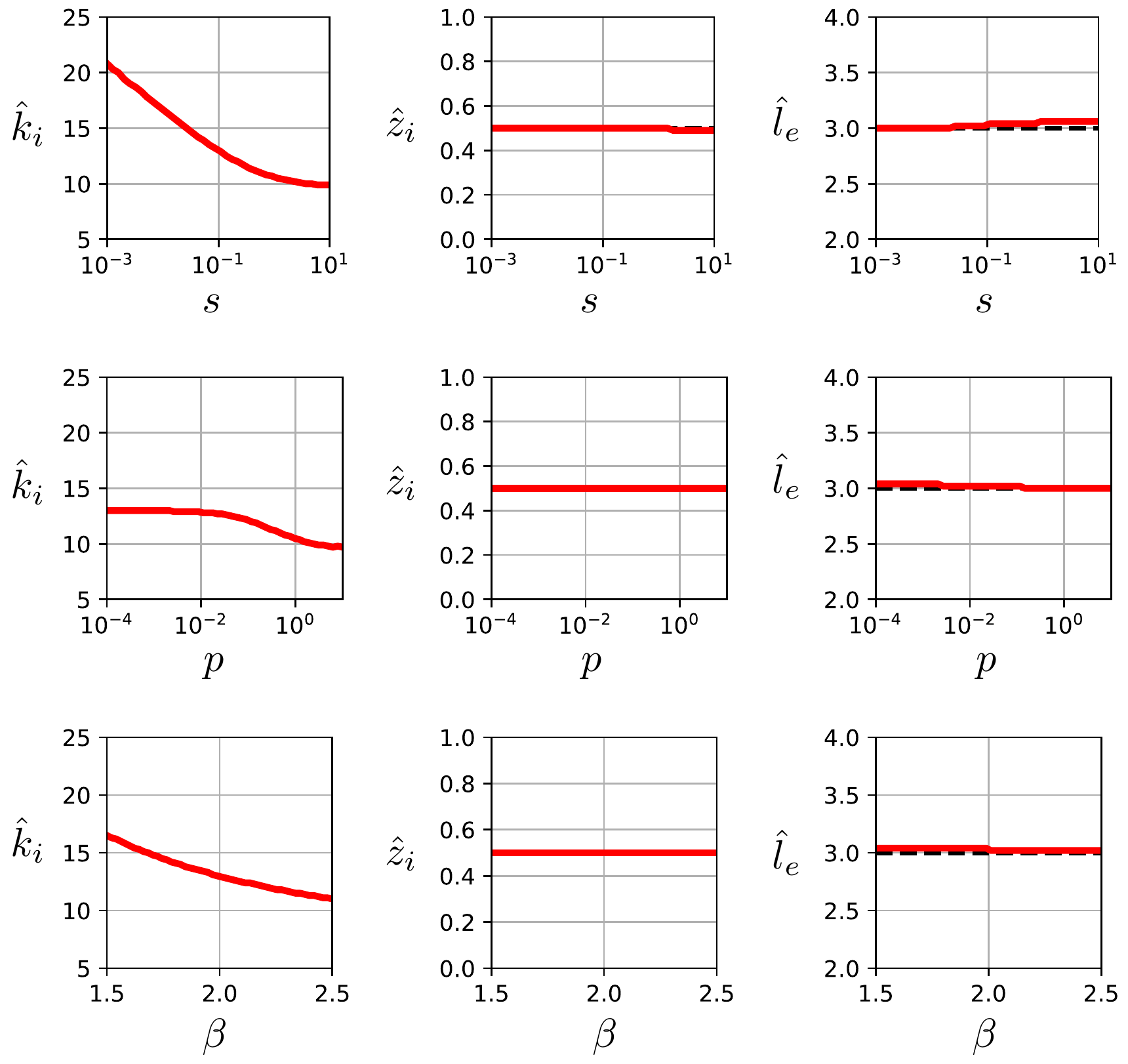}
\caption{Dependence of the optimal parameters $\hat k_i$ on the external parameters $s$, $p$ and $\beta$. The default parameters are as in Fig.~\ref{fig:localopt}. The solution (red line) is found to be given by Eq.~\eqref{eq:tt} (black dotted line, masked in most case under the red line), except for $\hat l_e$ where small deviations are visible. The dependence of $\hat k_e$ is not shown as $\hat k_e=\infty$ in all cases.\label{fig:optext}}
\end{center} 
\end{figure}

\begin{figure}[h]
\begin{center}
\includegraphics[width=.45\linewidth]{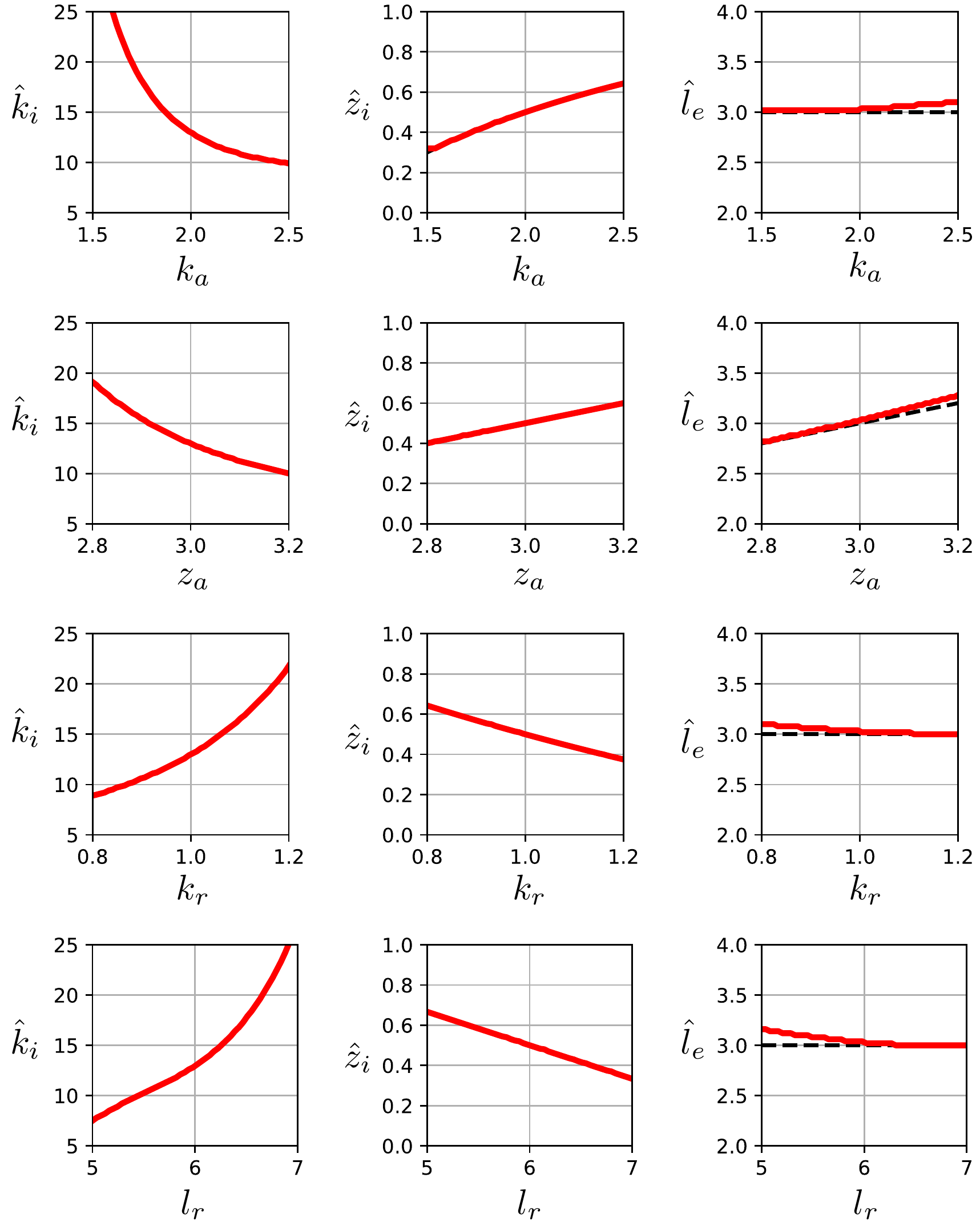}
\caption{Dependence of the optimal parameters $\hat k_i$ on the physical parameters of the substrate $k_a$, $z_a$, $k_r$ and $l_r$. The default parameters are as in Fig.~\ref{fig:localopt}. The solution (red line) is found to be given by Eq.~\eqref{eq:tt} (black dotted line, masked in most case under the red line), except for $\hat l_e$ where small deviations are visible. The dependence of $\hat k_e$ is not shown as $\hat k_e=\infty$ in all cases.\label{fig:optphys}}
\end{center} 
\end{figure}

\end{document}